\documentclass[preprints,article,accept,moreauthors,pdftex]{Definitions/mdpi} 
\firstpage{1} 
\pubvolume{xx}
\issuenum{1}
\articlenumber{5}
\pubyear{2019}
\copyrightyear{2019}
\history{Received: date; Accepted: date; Published: date}
\Title{Integrated Replay Spoofing-aware Text-independent Speaker Verification}

\Author{Hye-jin Shim $^{\dagger}$, Jee-weon Jung $^{\dagger}$, Ju-ho Kim  and Ha-jin Yu $^{}$*}

\AuthorNames{Hye-jin Shim, Jee-weon Jung, Ju-ho Kim and Ha-jin Yu}

\address{School of Computer Science, University of Seoul, Seoul, 02504, Republic of Korea; shimhz6.6@gmail.com}

\corres{Correspondence: hjyu@uos.ac.kr; }

\firstnote{These authors contributed equally to this work.} 
\abstract{
A number of studies have successfully developed speaker verification or presentation attack detection systems. 
However, studies integrating the two tasks remain in the preliminary stages.
In this paper, we propose two approaches for building an integrated system of speaker verification and presentation attack detection: an end-to-end monolithic approach and a back-end modular approach. 
The first approach simultaneously trains speaker identification, presentation attack detection, and the integrated system using multi-task learning using a common feature. 
However, through experiments, we hypothesize that the information required for performing speaker verification and presentation attack detection might differ because speaker verification systems try to remove device-specific information from speaker embeddings while presentation attack detection systems exploit such information. 
Therefore, we propose a back-end modular approach using a separate DNN for speaker verification and presentation attack detection. 
This approach has thee input components: two speaker embeddings (for enrollment and test each) and prediction of presentation attacks. 
Experiments are conducted using the ASVspoof 2017-v2 dataset, which includes official trials on the integration of speaker verification and presentation attack detection. 
The proposed back-end approach demonstrates a relative improvement of 21.77\% in terms of the equal error rate for integrated trials compared to a conventional speaker verification system.}

\begin{document}
\section{Introduction}
Recent advances in deep neural networks (DNNs) have improved the performance of speaker verification (SV) systems, including short-duration and far-field scenarios \cite{bhattacharya2019deep, jung2019rawnet, 9053871, jin2007far, 9004029}. 
However, SV systems are known to be vulnerable to various presentation attacks, such as replay attacks, voice conversion, and speech synthesis. 
These vulnerabilities have inspired research into presentation attack detection (PAD), which classifies given utterances as spoofed or not spoofed \cite{wu2015asvspoof, kinnunen2017asvspoof, todisco2019asvspoof} where many DNN-based systems have achieved promising results \cite{lai2019assert, jung2019replay, lavrentyeva2019stc}. 

Table \ref{tab:importance} demonstrates the vulnerability of conventional SV systems when faced with presentation attacks.
The performance is reported using the three types of equal error rates (EERs) described in Table \ref{tab:eer_types} \cite{todisco2018integrated}. 
Table \ref{tab:eer_types} shows the target and non-target trials for calculating the EER, which are represented by 1 and 0, respectively. 
Zero-effort (ZE)-EER describes the conventional SV performance without considering the presence of presentation attacks. 
PAD-EER denotes the EER for PAD which only considers whether an input is spoofed. 
Integrated speaker verification (ISV)-EER describes overall performance, considering both speaker identity and spoofing. 
We refer to ``replay spoofing-aware SV’’ as an ISV task and report its performance using ISV-EER. 
Results show that the EER of SV degrades to 33.72\% with replayed utterances; this fatal performance degradation supports the necessity of a spoofing-aware ISV %please define.
system.
In this paper, PAD refers to replay attacks, because the ASVspoof2017 dataset only focuses on replay attack detection which is known to be the easiest yet effective attack.
Therefore, the following three tasks are considered: SV, PAD, ISV, and performance is evaluated using ZE-EER, PAD-EER, and ISV-EER.

While a number of studies have worked to develop independent systems for SV and PAD, few have sought to integrate the SV and PAD systems \cite{sahidullah2016integrated, todisco2018integrated, sizov2015joint, dhanush2017factor, li2019multi, li2020joint}. 
More specifically, this handful of studies proposed approaches such as cascaded, parallel \cite{sahidullah2016integrated, todisco2018integrated}, and joint systems \cite{sizov2015joint, li2019multi, li2020joint}. 
Most existing studies used common features to integrate the two tasks for system efficiency.
Section \ref{sec:related} further takes up this existing body of work.

\begin{table}[t]
  \caption{Difference in equal error rate (EER) according to the existence of replay non-target trials. Results demonstrate the vulnerability of speaker verification systems that are unaware of PAD.}
  \label{tab:importance}
  \centering
  \begin{tabular}{lccc}
    \toprule
    & ZE-EER & PAD-EER & ISV-EER\\
    \midrule
    SV baseline & 9.58 & 33.72 & 19.98\\
    \bottomrule
  \end{tabular}
\end{table}
\begin{table}[t]
  \caption{Three types of EERs reported in this paper. Enrollment utterance is always bona-fide (i.e. genuine, not replayed). Target: enrollment and test utterances are uttered by an identical speaker and are bona-fide; ZE non-target: enrollment and test utterances are uttered by different speakers and are bona-fide; Replay non-target: enrollment and test utterances are uttered by an identical speaker and test utterance is replay spoofed.}
  \centering
  \label{tab:eer_types}
  \begin{tabular}{lccc}
    \toprule
    & \multirow{2}{*}{Target} & ZE & Replay\\
& & non-target & non-target \\
    \midrule
    ZE-EER & 1 & 0 &\\
    PAD-EER & 1 &  & 0\\
    ISV-EER & 1 & 0 & 0\\
    \bottomrule
  \end{tabular}
\end{table}

\begin{figure*}[t]
  \centering
  \includegraphics[width=\linewidth]{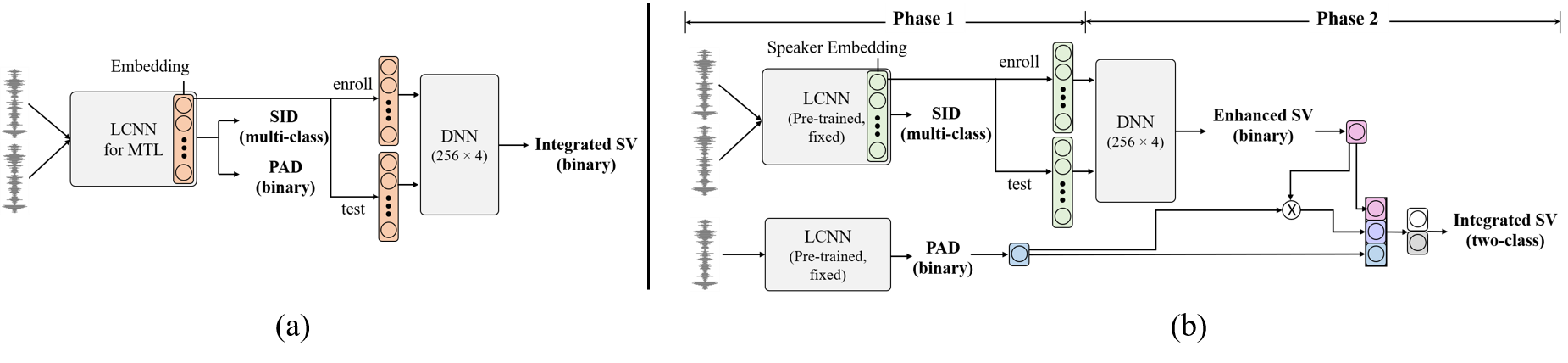}
  \caption{(a): An end-to-end architecture that trains embeddings (used for speaker identification (SID) and presentation attack detection (PAD)). LCNN and MTL refer to the light cnn and multi-task learning, respectively; (b): a separate architecture that inputs speaker embeddings from SID and PAD results and outputs the Integrated result of speaker verification (SV) and PAD.}
  \label{fig:int_architect}
\end{figure*}

In this paper, we propose two spoofing-aware frameworks for the ISV task, illustrated in Figure \ref{fig:int_architect}. 
We use a light cnn (LCNN) architecture \cite{LCNN} for both frameworks; this choice is based on its success in various PAD studies \cite{lavrentyeva2017audio, lavrentyeva2019stc}. 
The first proposed framework expands existing work by proposing a monolithic end-to-end (E2E) architecture. 
More specifically, it conducts speaker identification (SID) and PAD to train a common feature using multi-task learning (MTL) \cite{MultitaskLearning}.
Concurrently, it uses the embeddings to compose trials and conduct the ISV task. 
Using the sum of SID, PAD, and ISV losses, the entire DNN is jointly optimized. 
However, based on tendencies observed during internal experiments, we hypothesize that training a common feature for the ISV task may not be ideal because the properties required for each task differ: the PAD task representation uses device and channel information while SV needs to remove it (further discussed in Section \ref{sec:proposed}).

Based on our hypothesis, we propose a novel modular approach using a separate DNN. 
This approach inputs two speaker embeddings (for enrollment and test each) and a PAD prediction to make the ISV decision. 
It adopts a two-phase approach. 
In the first phase, the speaker identifier and PAD system are trained separately. 
In the second phase, speaker embeddings are extracted from a pre-trained speaker identifier \cite{d-vector}, and the embeddings and PAD prediction results are fed to a separate DNN module. 
Using this framework, we achieved a 21.77\% relative improvement in terms of ISV-EER (We use the trial in \cite{asv2017_url} for calculating ISV-EER).

The contributions of this paper are as follows:
\begin{enumerate}
    \item Propose a novel E2E framework that jointly optimizes SID, PAD, and the ISV task;
    \item Experimentally validate the hypothesis that the discriminative information required for the SV and the PAD task may be distinct, requiring separate front-end modeling;
    \item Propose a separate modular back-end DNN that takes speaker embeddings and PAD predictions as an input to make ISV decisions.
\end{enumerate}
The remainder of the paper is organized as follows. 
Section \ref{sec:related} details related work on the integrated system of SV and PAD.
Section \ref{sec:proposed} introduces the two proposed frameworks. 
Section \ref{sec:exp} presents our experiments and results and the paper is concluded in Section \ref{sec:conclusion}.

\section{Related work}
\label{sec:related}
In this section, we introduce the two studies most relevant to this study \cite{todisco2018integrated, li2019multi, li2020joint}. 
Firstly, Todisco et al. \cite{todisco2018integrated} propose a separate modeling of two Gaussian back-end systems with a unified threshold for both SV and PAD tasks. 
Their study explores various acoustic features to find which ones best simultaneously suited both tasks. 
As organizers of the ASVspoof challenges, official trials for the ISV task are released in this study. 
For our purposes, it is important to highlight that these trials include both ZE and replayed non-target, which we use throughout this paper. 
However, Todisco et al. \cite{todisco2018integrated} reported the average of two EERs, ZE-EER and PAD-EER, because they separately modeled two Gaussian mixture models for each task. 

Li et al. \cite{li2019multi, li2020joint} extended Todisco et al.’s work \cite{todisco2018integrated} by proposing an integrated ISV system; this study is the first that reports an ISV-EER. 
More specifically, they propose a three-phase training framework for extracting an embedding for the ISV task, followed by a probabilistic linear discriminant analysis (PLDA) back-end. 
In the first phase, a MTL \cite{MultitaskLearning} framework is employed to train a common embedding for both SV and PAD tasks. 
In the second and third phases, the embedding is adapted to fit the ISV task. 
However, because the DNN is adapted in the third phase to fit the enrollment speakers, it has limitations for real-world scenarios. 
In addition, because the performance is reported does not exploit organizer's official trials, it is difficult to compare the performance with the literature. 

In this paper, we first propose an E2E framework, illustrated in Figure \ref{fig:int_architect}(a), that extends the work of Li et al. \cite{li2019multi, li2020joint} in two aspects. 
First, we adopt a single phase training approach by using three loss functions for SID, PAD, and ISV. 
Second, our framework directly outputs a spoofing-aware score without using a separate back-end system. 

\section{Integrated speaker verification}
\label{sec:proposed}
In this section, we describe the two proposed  frameworks for conducting speaker verification that are aware of presentation attacks, as shown in Figure \ref{fig:int_architect}. 

\subsection{End-to-end monolithic approach}
We first propose an E2E monolithic approach.
This architecture simultaneously trains all components, including SID, PAD, and ISV, using a common feature, as illustrated in Figure \ref{fig:int_architect}(a). 
The loss function for training the proposed E2E architecture comprises three components: a categorical cross-entropy (CCE) loss for SID, a binary cross-entropy (BCE) loss for PAD, and a two-class BCE loss for ISV. 
When a mini-batch is input for training, the proposed system first conducts SID and PAD with an MTL framework. 
Then, it composes a number of trials. A trial consists of two embeddings, one for enroll and the other for test. 
The ISV prediction is made by feed-forwarding the two embeddings through a few fully-connected layers. 
The entire DNN is jointly optimized using the sum of three loss functions.
The objective function $Loss$ is defined as follows: 
  \begin{equation}
    \begin{aligned}
    Loss = Loss_{SID} + Loss_{PAD} + Loss_{ISV}
    \end{aligned}
  \end{equation}
where $Loss_{SID}$ refers to the CCE loss for SID, $Loss_{PAD}$ is the BCE loss for PAD, and $Loss_{ISV}$ denotes the CCE loss for ISV. 

However, we find consistent tendencies that make it difficult to extract a common representation, i.e., feature, for performing both SV and PAD tasks through experiments. 
Therefore, we hypothesize that, although SV and PAD tasks are closely related in the scenario, the discriminative information required for each task collides. 
Speaker embeddings for the SV task requires robustness to device and channel difference; meanwhile, representation for the PAD task uses such information \cite{shim2018replay}. 
Additionally, both bona-fide and replayed utterances include the same speaker information, making it a less discriminative factor for the PAD task; meanwhile, it is key information for the SV task. 
The study of Sahidullah et al. \cite{sahidullah2016integrated} supports our hypothesis, which states that the SV and PAD tasks should exist independently. 
To validate our hypothesis, we conduct experiments using separately trained SV and PAD systems and MTL-based systems. 
We further detail these elements in Section \ref{sec:exp}.3. 

\begin{figure}[t!]
  \centering
  \includegraphics[width=\linewidth]{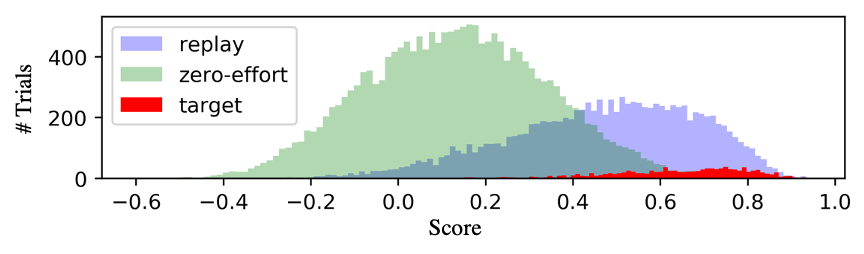}
  (a)
  \includegraphics[width=\linewidth]{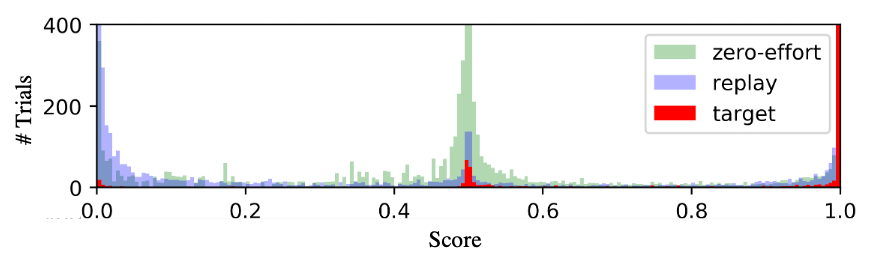}
  (b)
  \caption{Histograms of score distribution on the evaluation trials. (a): SV baseline where score is calculated using cosine similarity of two speaker embeddings. (b): The proposed modular system where three types of trials have three different distributions.}
  \label{fig:score_dist}
\end{figure}

\subsection{Back-end modular approach}
We also propose a novel modular approach using a separate DNN that takes speaker embeddings and PAD predictions as input to make an ISV decision.  
Figure \ref{fig:int_architect}(b) illustrates our second proposed system. 
Based on the hypothesis addressed in the previous subsection, we design an integrated system using a two-phase approach.
In the first phase, we separately train an SID system to extract speaker embeddings from the last hidden layer and a PAD system to extract a spoofing prediction. 
Then, we train the ISV system by using two speaker embeddings (one for enroll and the other for test) extracted from the SID system as a pair and a PAD label as an input. 
This system has an output layer with two nodes: the first node indicates ``acceptance’’ and the second node indicates ``rejection'’ for both ZE and replay trials. 

In Figure \ref{fig:int_architect}(b), the part trained in phase 2 is the proposed back-end ISV system.
It takes two speaker embeddings and multiplication of the two embeddings as input and a module of four fully-connected layers outputs a scalar that indicates whether they were uttered by the same speaker. 
The fully-connected layers comprise 256 nodes each and an output layer comprises one node with a sigmoid function.

Next, the SV and PAD prediction results and their multiplication are fed to a fully-connected layer to make the final decision. 
In an ideal scenario, the multiplication of the SV result and PAD prediction would indicate $1$ when both SV and PAD are positive and $0$ otherwise; we assume this multiplication would additionally inform the final decision. 
The objective function $Loss_{int}$ for the back-end modular approach comprises loss for the SV task and the loss for the final decision, defined as: 
  \begin{equation}
    \begin{aligned}
    Loss_{int} = \alpha \cdot Loss_{SV} + Loss_{ISV}
    \end{aligned}
  \end{equation}
where $Loss_{SV}$ and $Loss_{ISV}$ refer to the BCE loss of the SV task and the CCE loss of the ISV task, respectively, and $\alpha$ signifies the weight for the SV loss. 
We note that training the proposed back-end DNN with only $Loss_{ISV}$ results in overfitting. 

Based on a number of experiments that we omit here for the sake of brevity, we find two key components that make our proposed back-end DNN framework successful. 
First, we aim to model ZE and replayed trials into separate score distributions. 
Figures \ref{fig:score_dist}(a) and (b) respectively illustrate the score distributions of the evaluation trials of the SV baseline and the proposed modular back-end DNN. 
In Figure \ref{fig:score_dist}(a), the score refers to the cosine similarity of the two embeddings. 
Here, the score distribution of replay non-target trials severely overlaps with that of target trials. 
Since the existing speaker verification system does not consider the presentation attack detection, replay non-target and target trials can only be determined as targets because the replayed utterances and the bona-fide utterances share the same speaker information.
In our analysis, this resulted from embeddings that only considered speaker information in which replayed and bona-fide utterances coincided. 
In various experiments, it is impossible to model both replay and ZE non-target trials into the same score distribution. 
When one kind of non-target trial was successfully modeled, the other resulted in a distribution similar to uniform. 
Therefore, we aim to separate two non-target score distributions, specifically by modeling the score distribution of ZE non-target to have a mean of $0.5$ and replay the non-target to have a zero mean. 
To do so, we sequentially apply rectified linear unit (ReLU) and sigmoid activation functions to the output of SV before the last hidden layer for ISV. 
Figure \ref{fig:score_dist}(b) demonstrates the score distribution of the proposed method. 
The results demonstrat that three types of evaluation trials were modeled as intended (i.e., well generalized) in the case of evaluation trials, though these trials comprised unknown speakers and replay conditions. 

Second, we use actual PAD labels instead of PAD predictions of the spoofing DNN in the training phase. 
This is based on empirical comparisons in which the use of PAD predictions in the training phase worsened the performance. 
In our analysis, using PAD labels in the training phase was more helpful because even a small number of misclassified utterances among PAD predictions can interrupt the training of the proposed DNN. 
Notably, we empirically observed model collapse when training the proposed modular DNN using PAD predictions. 

\section{Experiments and results}
\label{sec:exp}
\subsection{Dataset}
\label{ssec:db}
All experiments in this study were conducted using the ASVspoof2017-v2 dataset \cite{delgado2018asvspoof}, because the official trials do not specify integrated systems for the ASVspoof2019 dataset.
Therefore, we evaluated the proposed integrated system by utilizing the official trials reported in \cite{todisco2018integrated} on the ASVspoof2017-v2 dataset.
We used training and development sets to train all systems comprising 2267 bona-fide and 2457 replay spoofed utterances from 18 speakers. 
To evaluate speaker verification and presentation attack detection performances, we measured the ZE-EER and the PAD-EER using the ASVspoof2017 joint PAD+SV evaluation trial. 
This trial comprised 1106 target, 18,624 ZE, and 10,878 replayed trials.  We used target and ZE for ZE-EER and target and replayed for PAD-EER evaluations. 

\subsection{Experimental configurations}
We used PyTorch, a Python deep learning library, for all experiments. 
For all DNNs, we input 64-dimensional Mel-filterbank features with utterance-level mean normalization following \cite{shim2018replay}. 
We applied weight decay with $\lambda=1e^{-4}$, and optimized with an AMSGrad optimizer \cite{reddi2019convergence}. 

Regarding our use of ASVspoof2017-v2, we found that relatively thin LCNN structures were helpful for performance improvement; this may have been a result of the small size of the dataset. 
In addition, we also found that minute changes to the DNN greatly influence the performance because of the small data scale; therefore, a relatively thin structure remained particularly helpful for performance improvement. 
To derive a value between 0 and 1 for the PAD task, we used a network architecture identical to that of \cite{lavrentyeva2019stc} but replaced the angular margin softmax activation \cite{wang2018cosface} with a sigmoid function.
We also modified the architecture for the SV task based on \cite{lavrentyeva2019stc}.
Speaker embeddings had a dimensionality of 1024.

\subsection{Results analysis}
\begin{table}[t]
    \caption{Results of various architectures using the proposed monolithic E2E framework for the ISV task. Numbers in bold represent the best results.}
  \label{tab:prop_e2e}
  \centering
  \begin{tabular}{lccc}
    \toprule
    System & ZE-EER (SV) & PAD-EER & ISV-EER\\
    \midrule
    \#1 & \textbf{18.52} & \textbf{15.73} & 18.44 \\
    \#2-SE & 18.99 & 15.90 & \textbf{17.90}\\
    \#3-split & 19.43 & 37.31 & 26.40\\
    \bottomrule
  \end{tabular}
\end{table}
\begin{table}[t]
  \caption{Experimental results showing that the required discriminative information differs for SV and PAD (Sid: speaker identification, PAD: presentation attack detection, Int: integrated speaker verification). Numbers in bold represent the best results.}
  \label{tab:unfit}
  \centering
  \begin{tabular}{lccc}
    \toprule
    Train loss & DNN arch & ZE-EER (SV) & PAD-EER\\
    \midrule
    Sid & SV & \textbf{9.58} & -\\
    Sid+PAD & SV & 17.53 & 13.69\\
   \midrule
   PAD & PAD & - & \textbf{10.60}\\
   PAD+Sid & PAD & 19.16 & 12.17\\
   \bottomrule
  \end{tabular}
\end{table}

Table \ref{tab:prop_e2e} describes the results of the proposed E2E framework with a monolithic approach. 
System \#1 refers to the proposed architecture that jointly optimizes SID, PAD, and ISV loss (see Figure \ref{fig:int_architect}(a)). 
System \#2-SE is the result of applying squeeze-excitation (SE) \cite{hu2018squeeze} based on its recent application to PAD \cite{lai2019assert}.
System \#3 describes the result of assigning three max feature map (MFM) blocks \cite{LCNN} for SID as well as for PAD after the first three MFM blocks. 
Because most of the system’s performance measures deteriorated compared to the SV baseline, we concluded that the monolithic E2E approach was not ideal for the ISV task. 
While the results of the experiments were different from what we expected, they nevertheless served as a springboard for establishing a new hypothesis.

Table \ref{tab:unfit} addresses the validation of our hypothesis in Section \ref{sec:proposed} that the discriminative information for the SV and the PAD task are distinct based on the results of Table \ref{tab:prop_e2e}. 
To validate our hypothesis, we trained our SV and PAD baselines with and without additional loss for extracting common embeddings. 
Here, the first and third rows refer to the SV and PAD baselines and the second and fourth rows refer to the usage of the MTL framework. 
The results demonstrated that, in both baselines, additionally adopting another loss function degraded performance.

\begin{table}[t]
  \caption{Results of the proposed modular approach for the ISV task. Numbers in bold represent the best results.}
  \label{tab:prop_modular}
  \centering
  \begin{tabular}{lccc}
    \toprule
    System & ZE-EER (SV) & PAD-EER & ISV-EER\\
    \midrule
    \#4-w/o mul & 20.52 & 19.77 & 20.48\\
    \#5-w mul & 15.59 & 18.06 & 16.66\\
    \midrule
    \#6-loss weight  & 15.22 & \textbf{14.55} & 15.91\\ 
    \#7-DNN arch & \textbf{14.32} & 15.46 &\textbf{15.63}\\
    \bottomrule
  \end{tabular}
\end{table}

\begin{table}[t]
  \caption{Comparison of the SV baseline, our proposed modular DNN, and other work using the official trials for the ISV task. Numbers in bold represent the best results.}
  \label{tab:compare_with_asvcm}
  \centering
  \begin{tabular}{lccc}
    \toprule
    & ZE-EER & PAD-EER & ISV-EER\\
    \midrule
    SV Baseline & 9.58 & 33.72 & 19.98\\
    \textbf{\#7-Ours} & 14.32 & \textbf{15.46} & \textbf{15.63}\\
    %\midrule
    %Todisco et al. \cite{todisco2018integrated} & \textbf{4.71} & 18.11 & -\\

    \bottomrule
  \end{tabular}
\end{table}

Table \ref{tab:prop_modular} summarizes the results of performance improvement across various attempts to improve the performance of the proposed method in the back-end modular approach.
The comparison of Systems \#4 and \#5 shows the effectiveness of using multiplication of the SV result and PAD prediction for the ISV task. 
System \#6 refers to the result of setting weights to the SV task in the training phase where we set the $\alpha$ to 20. 
System \#7 shows the result of reducing the number of nodes per hidden layer. 

Finally, Table \ref{tab:compare_with_asvcm} compares our proposed modular approach with the SV baseline and existing work \cite{todisco2018integrated} using official trials. 
The results demonstrated that the proposed approach stabilizes unbalanced performance between ZE-EER and PAD-EER. 
Compared with the SV baseline, which does not consider PAD attacks, we achieved a relative improvement of 21.77\%. 
It is important to note here is that we were unable to compare the ISV-EER with that of Todisco et al. \cite{todisco2018integrated}, although it is the only study that reported performance using official trials. 
Because it proposed a unified threshold for conducting SV and PAD tasks, ISV-EER results using the full trial do not exist. 

\section{Conclusion}
\label{sec:conclusion}
In this paper, we investigated the integration of speaker verification and presentation attack detection. 
We proposed two methods for their integration: an E2E monolithic approach and a back-end modular approach. 
The proposed E2E approach composes a single DNN that simultaneously trains SID, PAD, and ISV using a common feature. 
However, experimental results of the E2E approach led us to hypothesize that the discriminative information for SID and PAD differs; most configurations demonstrated performance degradation. 
Based on our hypothesis, we further proposed another framework using a separate back-end DNN that takes speaker embeddings and a PAD prediction extracted from pre-trained SV and PAD systems as input.
The effectiveness of our proposed systems was verified using official trials for the ISV task, where we achieved an EER of 15.63\%, showing 21.77\% relative improvement. 
Additionally, the proposed back-end modular system can be used with any SV and PAD systems because both of its inputs (i.e., speaker embeddings and PAD predictions) are extracted from pre-trained systems.
Thus, we expect that the proposed method will continue to enhance performance when improved speaker embeddings and PAD predictions from enhanced systems are input. 
Our future works include validating how well the proposed modular system's can generalize towards different speaker embeddings and PAD predictions.

\authorcontributions{Conceptualization, investigation, writing---original draft preparation and editing, H.-J.S., J.-W.J.; writing---review and editing, J.-H.K.; supervision, writing---review and editing, H.-J.Y. All authors have read and agreed to the published version of the manuscript.}

%%%%%%%%%%%%%%%%%%%%%%%%%%%%%%%%%%%%%%%%%%
\funding{This research was funded by the Ministry of Science, ICT and Future Planning, Grant number PA-J000001-2017-101. }

%%%%%%%%%%%%%%%%%%%%%%%%%%%%%%%%%%%%%%%%%%
\acknowledgments{This research was supported by Projects for Research and Development of Police Science and Technology under the Center for Research and Development of Police Science and Technology and the Korean National Police Agency funded by the Ministry of Science, ICT and Future Planning (Grant No. PA-J000001-2017-101)}

%%%%%%%%%%%%%%%%%%%%%%%%%%%%%%%%%%%%%%%%%%
\conflictsofinterest{The authors declare no conflict of interest.}

%%%%%%%%%%%%%%%%%%%%%%%%%%%%%%%%%%%%%%%%%%
% Citations and References in Supplementary files are permitted provided that they also appear in the reference list here. 

%=====================================
% References, variant A: internal bibliography
%=====================================
\reftitle{References}
%\begin{thebibliography}{999}
% Reference 1
%\bibitem[Author1(year)]{ref-journal}
%Author1, T. The title of the cited article. {\em Journal Abbreviation} {\bf 2008}, {\em 10}, 142--149.
% Reference 2
%\bibitem[Author2(year)]{ref-book}
%Author2, L. The title of the cited contribution. In {\em The Book Title}; Editor1, F., Editor2, A., Eds.; Publishing House: City, Country, 2007; pp. 32--58.
%\end{thebibliography}

% The following MDPI journals use author-date citation: Arts, Econometrics, Economies, Genealogy, Humanities, IJFS, JRFM, Laws, Religions, Risks, Social Sciences. For those journals, please follow the formatting guidelines on http://www.mdpi.com/authors/references
% To cite two works by the same author: \citeauthor{ref-journal-1a} (\citeyear{ref-journal-1a}, \citeyear{ref-journal-1b}). This produces: Whittaker (1967, 1975)
% To cite two works by the same author with specific pages: \citeauthor{ref-journal-3a} (\citeyear{ref-journal-3a}, p. 328; \citeyear{ref-journal-3b}, p.475). This produces: Wong (1999, p. 328; 2000, p. 475)

%=====================================
% References, variant B: external bibliography
%=====================================
%\externalbibliography{yes}
%\bibliography{your_external_BibTeX_file}

%%%%%%%%%%%%%%%%%%%%%%%%%%%%%%%%%%%%%%%%%%

%% for journal Sci
%\reviewreports{\\
%Reviewer 1 comments and authors’ response\\
%Reviewer 2 comments and authors’ response\\
%Reviewer 3 comments and authors’ response
%}

\newpage
\bibliography{mybib}

%%%%%%%%%%%%%%%%%%%%%%%%%%%%%%%%%%%%%%%%%%
\end{document}